\documentclass[useAMS,usenatbib]{mn2e}

\usepackage{graphicx}
\usepackage[fleqn]{amsmath}
\def\b#1{\bmath#1}
\def\d{{\rm d}}
\def\e{{\rm e}}
\def\E{{\cal E}}
\def\L{{\cal L}}
\def\Lgh{L_{\rm GH}}

\def\pr{{\rm p}}
\def\psf{{\rm psf}}

\def\vp{{v_{\rm p}}}
\long\def\crap#1{}
\def\=#1{{\langle#1\rangle}}

\title[Black hole masses]{Constraining black hole masses from stellar
  kinematics by summing over all possible distribution functions} \author[S. J.
Magorrian]{John Magorrian\thanks{E-mail:
    magog@thphys.ox.ac.uk}\\
  Rudolf Peierls Centre for Theoretical Physics, 1 Keble Road, Oxford
  OX1 3NP}
\begin{document}

\date{}


\maketitle

\label{firstpage}

\begin{abstract}
  When faced with the task of constraining a galaxy's potential given
  limited stellar kinematical information, what is the best way of
  treating the galaxy's unknown distribution function (DF)?  Using the
  example of estimating black hole (BH) masses, I argue that the
  correct approach is to consider all possible DFs for each trial
  potential, marginalizing the DF using an infinitely divisible prior.
  Alternative approaches, such as the widely used maximum penalized
  likelihood method, neglect the huge degeneracies inherent in the
  problem and simply identify a single, special DF for each trial
  potential.

  Using simulated observations of toy galaxies with realistic amounts
  of noise, I find that this marginalization procedure yields
  significantly tighter constraints on BH masses than the conventional
  maximum-likelihood method, although it does pose a computational
  challenge which might be solved with the development of a suitable
  algorithm for massively parallel machines.  I show that in practice
  the conventional maximum-likelihood method yields reliable BH
  masses with well-defined minima in their $\chi^2$ distributions,
  contrary to claims made by Valluri, Merritt \& Emsellem.

\end{abstract}


\section{Introduction}

A fundamental application of stellar dynamics is using observations of
a galaxy's kinematics to constrain its potential~$\psi(\b x)$.  The
galaxy is normally assumed to be collisionless and in a steady state,
so that the dynamics of any population of stars are completely
described by its phase space distribution function (DF), $f(\b x, \b
v)$, which is the probability density of finding a star in the small
volume of phase space around $(\b x,\b v)$.  By Jeans' theorem, this
DF can depend on $(\b x,\b v)$ only through the integrals of motion of
the (unknown) potential.

Most approaches to this task begin by considering the simpler problem
of constraining $f$ given the observed data and some trial $\psi$.
The infinite-dimensional DF is parametrized by a finite sum of delta
functions \citep[e.g.,][]{schw79} or a truncated basis function
expansion \citep[e.g.,][]{dejonghe89,saglia00}, and the DF parameters
are adjusted to optimize the fit to the observations.  If no set of
parameters yields an acceptable fit while simultaneously representing
a DF that is everywhere non negative, then the assumed potential can
be ruled out.

This basic idea can be refined further.  \citet{RT88} pointed out that
naive application of this method will yield unrealistically spiky DFs,
leading them to advocate the use of entropy (or something similar) as
a regularizer.  This idea of regularizing the resulting DFs was made
more explicit by \citet{merritt93}, who cast the problem as one of
finding the maximum penalized log-likelihood, ${\cal
  L}'\equiv-{1\over2}\chi^2+\lambda P[f]$, which allows a trade off
between goodness of fit, as measured by $\chi^2$, and smoothness, as
measured by the penalty function $P$.
The choice of penalty function and the value to use for the tradeoff
parameter~$\lambda$ are arbitrary and subjective.
Given a range of trial potentials, one finds the maximum penalized
likelihood for each and then uses normal statistical methods to make
statements about how well constrained the potential is.  This general
approach has become the method of choice in stellar-dynamical searches
for supermassive black holes (hereafter BHs) in galaxy centres, with
choices of penalty function ranging from the mean-square second
derivative of the DF \citep[e.g.,][]{vdm98,capp02} or entropy
\citep[e.g.,][]{g03,sil05} through to models in which no regularization
whatsoever has been applied \citep[e.g.,][]{vdm98,ryan06}.  I refer to
these as ``maximum-likelihood'' or ``maximum-penalized likelihood''
methods.

Another way to look at the problem of constraining potentials is to
treat it as a straightforward mathematical inverse problem.
\citet{dm92} considered the case of constraining the potential of a
spherical galaxy given perfect knowledge of its projected DF in the
form of its line-of-sight velocity profiles (hereafter VPs)
$\L(R;\vp)$.  By using a set of higher-order Jeans equations they
showed that, given the potential, the DF $f(\E,J^2)$ is completely
determined by its projected VPs.  Constraining the potential, however,
proved less amenable to their methods, but they found that choosing a
potential too far from the true one would yield moments (and therefore
DFs) that became negative, ruling out that potential.

Of course, one never has perfect knowledge of the full projected DF.
More recently, \citet[][hereafter VME04]{vme04} investigated the
slightly less idealized problem of constraining the BH mass in a toy
axisymmetric galaxy given noiseless measurements of a restricted
number of (modified) moments of its VPs averaged over a number of
spatial bins on the sky.  They showed that even when the potential has
just one free parameter -- the BH mass -- there are many different
potentials that can fit the available kinematics almost perfectly,
even when the central spatial resolution of the kinematics is much
finer than the BH's sphere of influence.

One thing that all these methods have in common is that they consider
just one DF for each trial potential.  This is fine for the idealized
case considered by \citet{dm92}: given
perfectly resolved, noiseless projected VPs of a spherical galaxy
there is a unique DF for any assumed potential, even though this DF
may not be non-negative everywhere.  But when the available data have
finite spatial and velocity resolution, there will in general be many
perfectly sensible, non-negative DFs that yield equally good fits to
the data, and even more DFs producing fits that are only slightly
worse.  Intuitively, one might expect that the more such DFs a
potential admits, the more likely it is.

The purpose of the present paper is to revisit the the problem of
constraining BH masses from a thoroughly Bayesian perspective, showing
how it naturally incorporates this intuitive notion of counting up
DFs.  To illustrate the ideas, I use simulated observations of some
idealized spherical toy galaxies described in Section~2, modelling
them under the assumptions given in section~3.  In section~4 I test
how well the conventional maximum likelihood method recovers BH
masses and counter some of the more pessimistic conclusions of VME04.
Section~5 presents a Bayesian approach to the problem, which overcomes
some of the inconsistencies of the maximum likelihood method.
Finally, section~6 sums up and discusses the implications for BH
masses in real galaxies.

\section{Toy galaxies}
\label{sec:toy}

\subsection{Intrinsic properties}

My toy galaxies are spherical with luminosity density profile
\citep{dehnen93,tremaine94}
\begin{equation}
\label{eq:lumprof}
j(r) = {(3-\alpha)L\over4\pi}{a\over r^\alpha(a+r)^{4-\alpha}},
\end{equation}
and constant mass-to-light ratio~$\Upsilon$ for radii $r>0$, so that
the total stellar mass $M_\star = \Upsilon L$.  At $r=0$ there is a BH
of mass $M_\bullet=2\times10^{-3}M_\star$.  The galaxies used in this
paper all have inner density slope $\alpha=1.5$, for which the BH
dominates the kinematics inside a radus $0.015a$. 

By Jeans' theorem \citep{bt}, a spherical galaxy can be in equilibrium
only if its phase-space distribution function (DF) depends on $(\b
x,\b v)$ only through the integrals of motion $\E$ and~$\b J$, the
energy and angular momentum per unit mass.  The DFs of the toy
galaxies have the form \citep{cudd91}
\begin{equation}
f(\E,J^2) = J^{-2\beta} g(\E),
\label{eq:DF}
\end{equation}
where the parameter $\beta$ controls the degree of anisotropy, with
$\beta=1-\sigma_\phi^2/\sigma_r^2$.  I solve for $g(\E)$ given $j(r)$
and $M_\bullet$ using the method described in \citet{mt99}, and
present results for both isotropic ($\beta=0$) and mildly radially
anisotropic ($\beta=0.3$) toy galaxies.

\subsection{Observables}

The standard ``observations'' of each toy galaxy consist of its
luminosity-weighted VPs averaged over abutting shells, with 5 shells
per decade in radius whose centres run from $R_{\rm min}=10^{-3}a$ to
$R_{\rm max}=4a$.  These observations both resolve the
BH's sphere of influence and extend to more than twice the galaxies'
effective radii.  I calculate VPs~$\L(R;\vp)$ using the procedure
described in \citet{vdm00} and parametrize each using a Gauss--Hermite
series \citep{g93,vdMF93},
\begin{align}
   \label{eq:GHseries}
\Lgh(R;\vp) &= {\gamma\over\sqrt{2\pi}\sigma}
  \exp\left[-{1\over2}\left(v-V\over\sigma\right)^2\right]
\sum_{i=0}^\infty h_iH_i\left(v-V\over\sigma\right).
\end{align}
This expresses the VP as an underlying Gaussian with normalization
$\gamma$, mean~$V$ and dispersion~$\sigma$, modified by a sum of
Hermite polynomials~$H_i$.  For any reasonable choice of
$(\gamma,V,\sigma)$ it is straightforward to show that choosing
\begin{equation}
\label{eq:GHcoeff}
h_i = {1\over\sqrt2\gamma}
\int_{-\infty}^\infty \exp\left[-{1\over2}\left(v-V\over\sigma\right)^2\right]
L(v)H_i\left(v-V\over\sigma\right)\,{\rm d}v
\end{equation}
minimizes the mean-square deviation between $\L(R;\vp)$ and
$\Lgh(R;\vp)$.  Therefore, the $h_i$ are simply {\it modified moments}
of $\L(R;\vp)$.  I choose $(\gamma,V,\sigma)$ to be the parameters of
the best-fitting Gaussian to $\L(R;\vp)$, in which case $V$ and the
odd $h_i$ are zero, $(h_0,h_2)=(1,0)$ (by equs.\ \ref{eq:GHseries} and
\ref{eq:GHcoeff} above) and $h_4$ measures the lowest-order departure
of the VP from Gaussianity.
\begin{figure}
\begin{center}\includegraphics[width=0.8\hsize]{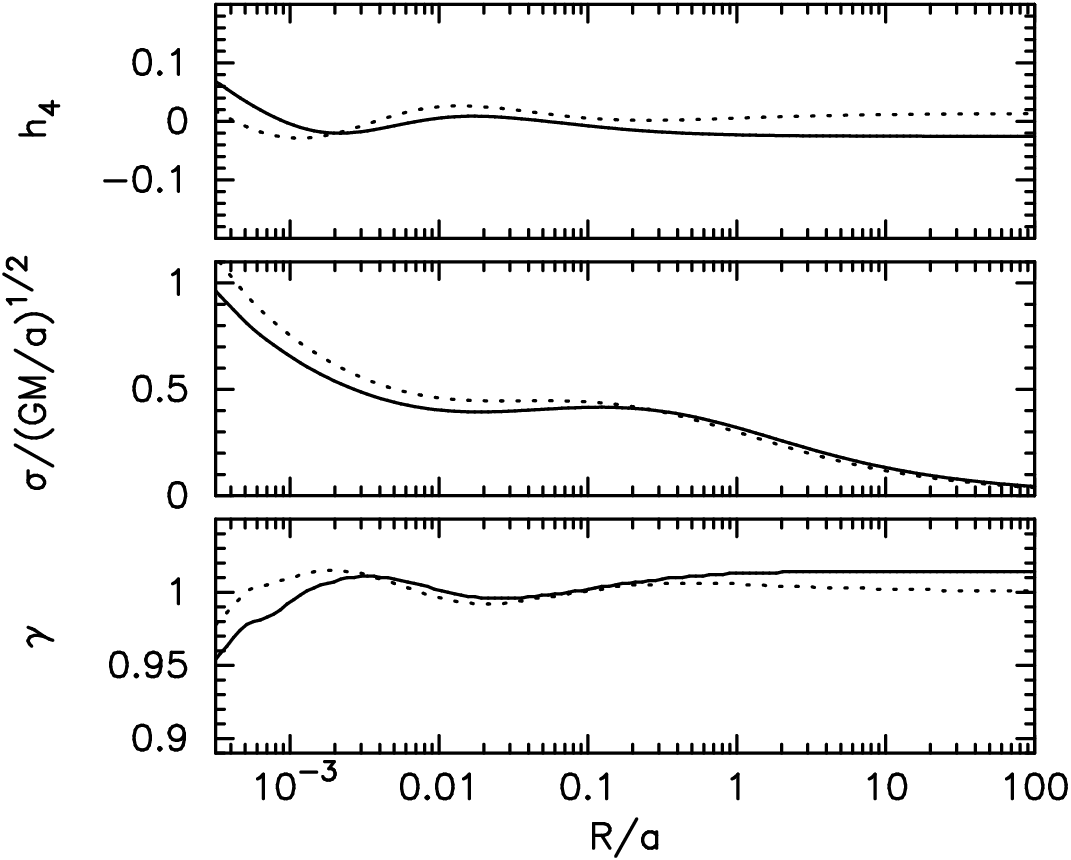}\end{center}
\caption{Gauss--Hermite coefficients of the line-of-sight VPs of the
  isotropic (solid curves) and anisotropic (dotted) toy galaxies.}
\end{figure}

Each realization of a toy galaxy then consists of 19 VPs.  I expand
each VP about its underlying best-fit Gaussian $(\gamma,V,\sigma)$,
and reduce the VP to four ``measurements'': the mean surface
brightness $I$, and the three lowest-order luminosity-weighted
modified moments ($Ih_0,Ih_2,Ih_4)$.  To these I add independent,
normally distributed errors $\Delta I=10^{-3}I$, $(\Delta h_0,\Delta
h_2,\Delta h_4) = (0.02, 0.05, 0.05)$, comparable to the formal errors
from observations of real galaxies.  Notice that the parameters of the
underlying Gaussian $(\gamma,V,\sigma)$ are {\it chosen} by me, not
measured, and so have no measurement uncertainties.

Some further comments on the use of the modified
moments~(\ref{eq:GHcoeff}) are in order.  \citet{ryan06} have shown
that Gauss--Hermite expansions are not particularly well suited for
parametrizing the VPs of real galaxy centres: real VPs can be very
strongly non-Gaussian, and in practice measurements of the
coefficients~$h_i$ are not independent, even for fixed
$(\gamma,V,\sigma)$.
I nevertheless use the Gauss--Hermite parametrization in the present
paper for the following reasons: truncated fourth-order Gauss--Hermite
expansions turn out to provide reasonably accurate fits to the VPs of
the toy galaxies; real observations have finite velocity resolution,
which is mimicked, at least qualitatively, by truncating the infinite
Gauss--Hermite expansion (e.g., compare the eigen-VPs in fig.~6 of
\citet{ryan06} with the Gauss--Hermite basis in fig.~1 of
\citet{vdMF93}); finally, using Gauss--Hermite expansions permits a
more direct comparison with VME04's method and results.


\section{Models}

The general modelling scheme is the same as that used in
\citet{ryan06}.  The model potentials~$\psi(r)$ have two free
parameters, the BH mass~$M_\bullet$ and the mass-to-light
ratio~$\Upsilon$, corresponding to mass densities of the form
\begin{equation}
  \label{eq:massprof}
  \rho(r) = {M_\bullet\over4\pi} \delta(r) + \Upsilon j(r),
\end{equation}
where $j(r)$ is given by~(\ref{eq:lumprof}).  Since the kinematics of
the toy galaxies are averaged over abutting shells and extend to many
effective radii, it turns out that $\Upsilon$ is very well constrained
by virtue of the virial theorem \citep{RT88}.  So, for the
results presented in this paper I simply fix $\Upsilon$ at its correct
value.  This means that model and galaxy potentials differ only in
their BH masses.

Having the potential~$\psi$, I discretize the DF
\begin{equation}
f(\E,J^2) = \sum_{i=1}^{n_\E}\sum_{j=1}^{n_J} f_{ij}
\delta(\E-\E_i)\delta(J^2-J^2_{ij}),
\label{eq:DFdoublesum}
\end{equation}
on an $n_\E\times n_J$ regular grid in phase space.  The points $\E_i$
are chosen through $\E_i=\psi(r_i)$ with the $r_i$ spaced
logarithmically between $10^{-5}a$ and $10^{3}a$.  There are $n_J$
values of angular momentum for each~$\E_i$, with $J^2_{ij}$ running
linearly between $0$ and $J^2_{\rm c}(\E_i)$, the angular momentum of
a circular orbit of energy~$\E_i$.  To avoid a rash of indices I
henceforth write the double sum~(\ref{eq:DFdoublesum}) as a single sum
over $n\equiv n_\E\times n_J$ points:
\begin{equation}
f(\E,J^2) = \sum_{i=1}^n f_i \delta(\E-\E_i)\delta(J^2-J^2_i).
\label{eq:DFdiscretization}
\end{equation}
This discretization effectively partitions phase space into abutting
rectangular cells, with the luminosity contained in each cell being
given by
\begin{equation}
 \label{eq:lumcell}
  L_i \equiv f_i\int_{V_i} g(\E,J^2)\,\d\E\d J^2,
\end{equation}
where $g(\E,J^2)$ is the density of states for the potential~$\psi$
and $V_i$ the volume occupied by the cell.  In section~\ref{sec:priors}
it will provide convenient to use a dimensionless luminosity
\begin{equation}
  \label{eq:flux}
  F_i\equiv {L_i\over L_{\rm s}},
\end{equation}
where $L_{\rm s}$ is a characteristic luminosity scale.

The models' projected observables $I(R)$,
$Ih_i(R)$ depend linearly on the orbit weights $f_i$, so that the
$\chi^2$ of a model with DF $\b f\equiv (f_1,\ldots,f_n)^T$ is the
quadratic form
\begin{equation}
  \label{eq:chisq}
\chi^2(\b f|\psi) = \left[{\b Q}-P(\psi)\cdot\b f\right]^T\cdot
\left[{\b Q}-P(\psi)\cdot\b f\right],
\end{equation}
where 
\begin{equation}
 {\b Q}\equiv \left( {\phantom{\Delta}I(R_1)\over\Delta I(R_1)}, 
                      {\phantom{\Delta}Ih_0(R_1)\over\Delta Ih_0(R_1)},
                         \cdots,
                      {\phantom{\Delta}Ih_4(R_N)\over\Delta Ih_4(R_N)}
               \right)^T
\end{equation}
is a column vector containing the list of observations, normalized by
their uncertainties, and  $P(\psi)$ is a projection matrix whose $n$
columns contain the contribution each DF component makes to the
model's prediction for $\b Q$.  The calculation of $P(\psi)$ is
described in Appendix~A.

\section{Fitting models to observations: the
  maximum-likelihood method}
\label{sec:maxlik}
Having a set of observations of a toy galaxy, let us first test how
well a simplified version of the standard maximum-likelihood
method \citep[e.g.,][]{vdm98,g03,vme04,ryan06} reproduces
the correct BH mass and its uncertainties.  The procedure is as
follows:
\begin{enumerate}
\item choose a trial BH mass $M_\bullet$ and calculate the
  corresponding potential~$\psi$;
\item calculate the projection matrices $P(\psi)$ appearing
  in~(\ref{eq:chisq});
\item use a non-negative least-squares algorithm \citep{lawsonhanson}
  to find $\chi^2_{\rm min}(\psi)$, the minimum value
  of~(\ref{eq:chisq}) subject to the constraint that all $f_i\ge0$;
\item assign a likelihood $\exp[-{1\over2}\chi^2_{\rm min}(\psi)]$ to
  the potential~$\psi$.
\end{enumerate}
One obtains constraints on $M_\bullet$ by considering a range
of~$M_\bullet$ and comparing their relative likelihoods.  To keep the
interpretation of the results as simple as possible, I do not impose
any regularization on the~$f_i$.  Section~\ref{sec:reg} below
discusses this further.

\begin{figure}
\begin{center}\includegraphics[width=0.8\hsize]{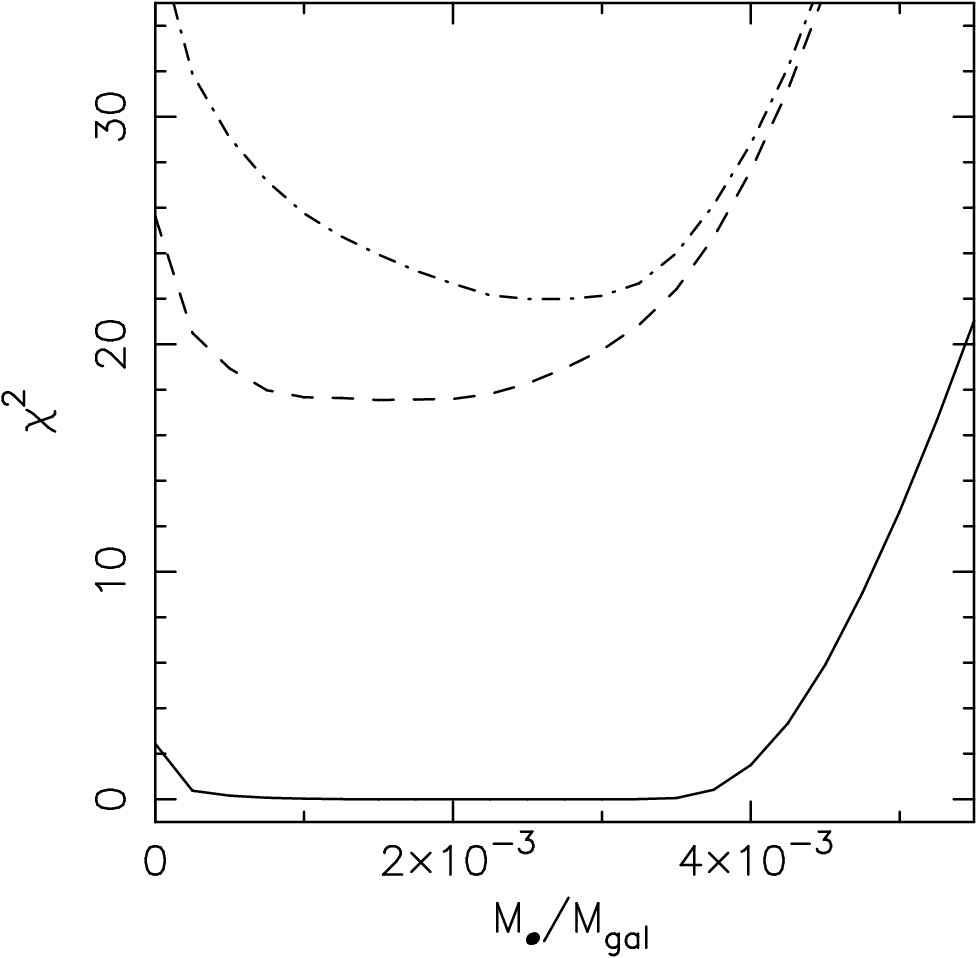}\end{center}
\caption{$\chi^2$ distributions returned by the conventional
  maximum-likelihood method (section~\ref{sec:maxlik}) for noiseless
  observations of an isotropic toy galaxy (solid curve) and for two
  different realizations of observations with simulated noise (dashed
  and dot-dashed curves).  }
\label{fig:chisqconven}
\end{figure}

\subsection{VME04's flat-bottomed $\chi^2(M_\bullet)$
  distributions}

One of the more alarming conclusions reached by VME04 was that the
problem of constraining BH masses is inherently strongly degenerate;
they found that a wide range of BH masses could provide equally good
fits to mock kinematics with realistic spatial resolution.
The main goal of this section of the present paper is to understand
this result and to investigate whether its implications really are as
negative as VME04 suggest.

The solid curve in figure~\ref{fig:chisqconven} plots
$\chi^2(M_\bullet)$ obtained for models with $n_\E\times
n_J=100\times10$ DF components when applied to noise-free observations
of the isotropic toy galaxy.  It demonstrates that VME04's central
result also holds for the simpler spherical case considered here: a
wide range of $M_\bullet$ can produce perfect fits to perfect
noiseless data.  For BH masses in the range $1.8\times10^{-3}\la
M_\bullet/M_{\rm gal}\la 3.0\times10^{-3}$, $\chi^2$ is of order
$10^{-25}$, rising to $\sim10^{-5}$ for $M_\bullet/M_{\rm
  gal}=1.7\times10^{-3}$ or $3.1\times10^{-3}$.  A model with {\it no}
BH can produce kinematics that differ by only a small amount
($\Delta\chi^2=2.4$) from the toy galaxy's.  Of course, these values
of $\chi^2$ are statistically meaningless since they do not account
for the fact that the observations in this contrived situation have
zero uncertainty; the increase of $\chi^2$ to $10^{-5}$
from its minimum value of $10^{-25}$ (which is zero to machine
precision) is actually very significant.

The other two curves in figure~\ref{fig:chisqconven} show the results
of adding two different realizations of noise to the simulated
dataset.  This makes $\chi^2(M_\bullet)$ become nicely rounded,
similar to what one finds in models of real galaxies
\citep[e.g.,][]{vdm98,g03}.  VME04, however, only presented results
for the noiseless case.

These results can be explained by remembering that, for a fixed potential,
$\chi^2$ is a quadratic form~(\ref{eq:chisq}) in the orbit weights~$\b
f$.
Since the number of unknowns is very much less than the number of
observations, this quadratic form is hugely degenerate: it resembles
more a multi-dimensional trough than a parabola.
If we relax the constraint that all $f_i\ge0$, then it turns out that
for all the models considered here -- independent of the value
of~$M_\bullet$ -- the value of $\chi^2$ at the bottom of the trough is
zero (to machine precision); the non-negativity constraint is
essential for constraining the BH mass.  Of course, for the correct
model with $M_\bullet=2\times10^{-3}$, the bottom of the trough passes
through the discretized version of the true DF~(\ref{eq:DF}), which is
well away from the boundaries given by $f_i\ge0$.
Then making a small change in the trial~$M_\bullet$ leads to a small
change in the projection matrix, particularly for those $f_i$
corresponding to the most tightly bound orbits, and therefore changes
the shape of the quadratic form slightly as well as the location of
its minimum.  Changing the potential too much moves the location of
the minimum to a region where at least one of the weights becomes
negative, so that the minimum value of $\chi^2$ in the subvolume
$f_i\ge0$ is no longer zero.

Adding noise simply shifts the centre of the quadratic form, with no
change in its shape.  For realistic amounts of noise, the centre is
shifted well into the region where many of the orbit weights are
negative, leading to the rounded $\chi^2(M_\bullet)$ profiles.

\subsection{The effects of signal to noise on the
uncertainties on $M_\bullet$}

In order to examine this more quantitatively, let us consider how the
uncertainties on $M_\bullet$ depend on the signal-to-noise ratio of
the simulated data.  To do this, we need a method of quantifying the
uncertainty on $M_\bullet$.  The accepted practice in this field is to
assume that the $\Delta \chi^2=1$ boundaries of $\chi^2_{\rm
  min}(M_\bullet)$ give reliable indicators of the 68 percent
confidence limits on $M_\bullet$ \citep[e.g.,][]{vdm98}.  This is
based on the assumption \citep[e.g.,][]{nr} that $\chi^2_{\rm
  min}(M_\bullet)$ is close to quadratic and therefore that the
probability distribution $\exp(-\chi_{\rm min}^2/2)$ is almost
Gaussian, but the results above and in VME04 show that this
assumption can be far from the truth.  So,
throughout this paper I use the mean and variance,
\begin{equation}
\begin{split}
\overline{M_{\bullet}} &\equiv A\int
 M_\bullet \exp\left[-{1\over2}\chi^2_{\rm min}(M_\bullet)\right]\d M_\bullet\\
\left(\Delta {M_{\bullet}}\right)^2& \equiv A\int (M_\bullet-\overline{M_\bullet})^2
 \exp\left[-{1\over2}\chi^2_{\rm min}(M_\bullet)\right]\d M_\bullet,
\end{split}
\label{eq:meanvar}
\end{equation}
to quantify the best-fitting $M_\bullet$ and its associated
uncertainty, the quantity $A$ being chosen to make
$A\int\exp[-\chi^2_{\rm min}]\,\d M_\bullet=1$.  I find that this
$\Delta M_\bullet$ agrees well with the (correctly calculated) 68
percent confidence intervals for the following.

\begin{figure}
\begin{center}\includegraphics[width=0.8\hsize]{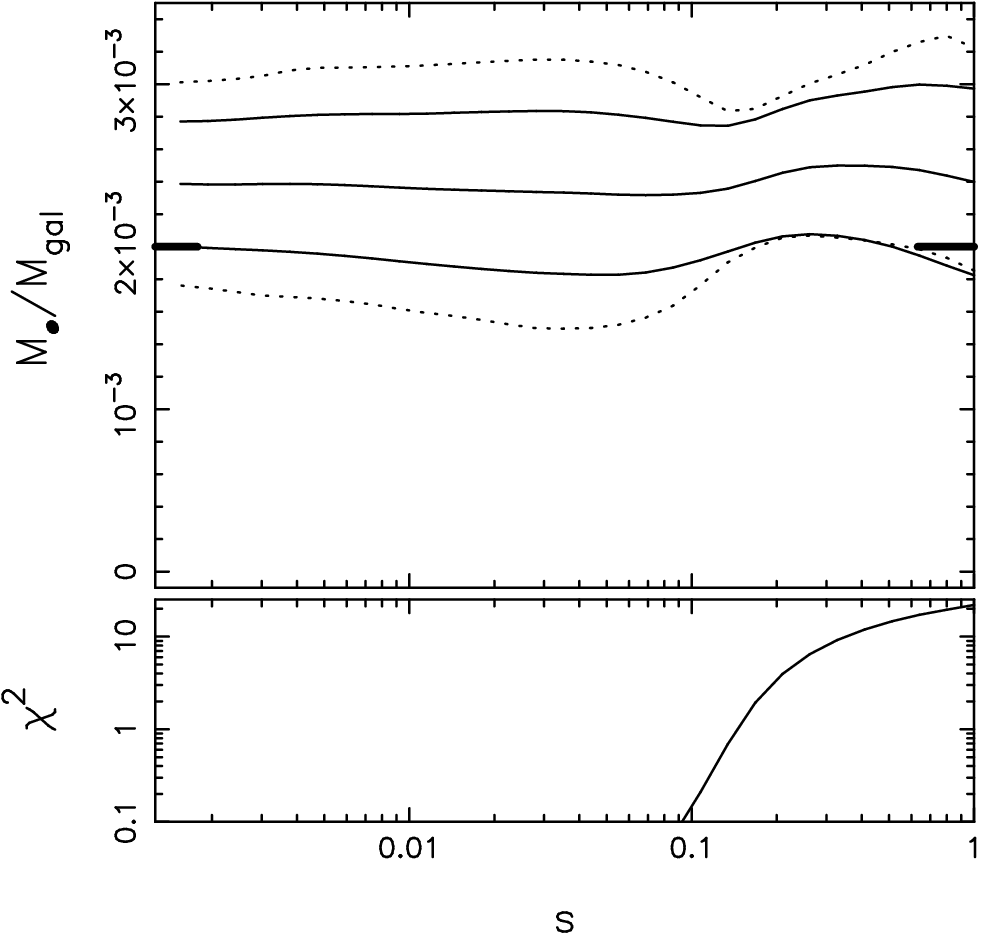}\end{center}
\caption{Effect of relative signal-to-noise ratio, $1/s$, on the BH
  mass $M_\bullet$ returned by the maximum-likelihood method for a
  typical realization of the noise in the observations.  The solid
  curves in the top panel plots the mean $M_\bullet$ and its formal
  uncertainty (eq.~\ref{eq:meanvar}) as a function of $s$, the
  relative size of the observational errors.  For comparison, the
  dotted curves show the uncertainties on $M_\bullet$ returned by the
  widely used $\Delta\chi^2=1$ criterion.  The bottom panel plots
  the corresponding minimum value of $\chi^2$.  The $s=1$ case
  corresponds to the dot-dashed curve in
  figure~\ref{fig:chisqconven}.}
\label{fig:varynoise}
\end{figure}

Figure~\ref{fig:varynoise} shows how $\overline{M_\bullet}$ and
$\Delta M_\bullet$ vary as one changes the size of the observational
uncertainties for one particular noise realization $\Delta\b Q$,
the errors $(\Delta I/I,\Delta h_0,\Delta h_2,\Delta h_4)=s\Delta\b Q$ being
shrunk by a factor~$s$ with respect to the ``standard'' observational
errors.  The following points hold for typical noise realizations
$\Delta\b Q$:
\begin{enumerate}
\item Given noiseless data ($s=0$), a range of $M_\bullet$ spanning
  $1.2\times10^{-3}M_{\rm gal}$ can produce perfect fits.  This is
  essentially the
  ``$\Delta\chi^2=1$'' measure of the uncertainty on $M_\bullet$
  applied to a uniform distribution.
\item The variance of a distribution uniform for $x\in [a,b]$ is given
  by $(b-a)^2/12$.  Therefore, the more useful estimate of the
  uncertainty given by equ.~(\ref{eq:meanvar}) is a factor $\sqrt{12}$
  smaller, or $0.35\times10^{-3}\,M_{\rm gal}$.
\item One can obtain perfect fits to the data for any $s\la0.1$, the
  precise upper bound on~$s$ depending on the particular noise realization.
\item Around the value of $s$ where the best-fitting $\chi^2$ starts
  to lift off from zero, the uncertainty $\Delta M_\bullet$ {\it
    drops} as $s$ increases.
\item Overall, $\Delta M_\bullet$ grows slowly with $s$; it grows
  by only a factor $\sim1.5$ between the noise-free $s=0$ and the more
  realistic $s=1$ situation.
\end{enumerate}

\begin{table}
\caption{BH mass estimates for toy galaxies using the conventional
  maximum-likelihood method }
 \begin{tabular}{cccccc}
$\beta$ & $n_{\E}\times n_J$ & $R_{\rm min}/a$ &
 $\langle{M_\bullet}/10^{-3}M_{\rm gal}\rangle$ & 
 $\langle{\Delta (M_\bullet/10^{-3}M_{\rm gal})^2}\rangle^{1/2}$ \\
\hline
0     & $100\times10$ & $10^{-3}$      & 2.08 & 0.81 \\
0     & $200\times20$ & $10^{-3}$      & 2.10 & 0.85 \\
0     & $100\times10$ & $10^{-4}$      & 1.89 & 0.34 \\
\noalign{\smallskip}
0.3   & $100\times10$ & $10^{-3}$      & 2.57 & 0.99 \\
0.3   & $200\times20$ & $10^{-3}$      & 2.55 & 1.06 \\
0.3   & $100\times10$ & $10^{-4}$      & 2.08 & 0.42 \\
  \hline
 \end{tabular}
 \medskip

 Columns are: the galaxy's anisotropy ($\beta$), the number of orbits used in the modelling
 ($n_{\E}\times n_J$) and the innermost extent of the projected
 kinematics ($R_{\rm min}$); the mean BH mass from many realizations
 $\langle M_\bullet\rangle$ and the typical formal uncertainty in $M_\bullet$.
\label{tab:conventional}
\end{table}

At first sight the last point might seem to suggest that there is
little point in obtaining very high signal-to-noise observations, but
of course one could extract higher-order information on the VPs from
such observations, such as $h_6$ or its equivalent, and, in some
cases, one could also use finer spatial binning.  Both of these would
act to reduce the degeneracy in $M_\bullet$ for $s=0$.

Having examined the effects of observational uncertainties on the
uncertainty on $M_\bullet$, let us now turn to the simpler question of
whether the maximum-likelihood method yields estimates of $M_\bullet$
close to the true BH mass.  Taking the $s=1$ situation of realistic
noise and averaging over many hypothetical datasets, the typical
formal error in $M_\bullet$ is about $0.8\times10^{-3}M_{\rm gal}$
(isotropic galaxy) or $1\times10^{-3}M_{\rm gal}$ (anisotropic
galaxy), both increasing only slightly as the number of DF components
used increases (Table~1).  The maximum-likelihood method yields fairly
strongly biased estimates of $M_\bullet$ for the anisotropic galaxy,
which nevertheless are well within the formal uncertainties.

\subsection{The appropriateness of regularization}
\label{sec:reg}

Apart from the strange dependence of the formal uncertainty $\Delta
M_\bullet$ on the signal-to-noise ratio, perhaps the most telling
feature of models obtained using the maximum-likelihood method is how
well they fit: they are too good to be true.  While adding realistic
amounts of noise to the observations removes the flat bottom in
$\chi^2$, the value of $\chi^2$ at the minimum remains very much less
than the number (76) of observed data points (e.g.,
figure~\ref{fig:chisqconven}).
These fits are implausibly good; the chances are tiny that the actual
values of $(I,h_0,h_2,h_4)$ in the real galaxy are all so close to the
observed estimates.  Furthermore, the models achieve this level of fit
by having only $\sim70$ of the $f_i$ greater than zero: the internal
kinematics of the model are very irregular.  It is important then to
consider how $M_\bullet$ is affected when one includes models that
yield more plausible fits to the data.

Following~\citet{merritt93}, the approach advocated by VME04 and
subsequently by \citet{cretem} is to regularize the DF, finding for
each~$M_\bullet$ the $\{f_i\}$ that maximize a penalized
log-likelihood, $-{1\over2}\chi^2+\lambda P[f]$.  The penalty function
$P[f]$ provides some arbitrary measure of the smoothness of the DF and
the parameter $\lambda$ is set by how much one is willing to trade off
goodness-of-fit for a smoother DF.  Although beguiling, this approach
is only marginally better than the conventional maximal likelihood
method used above, because both
\begin{enumerate}
\item identify a single privileged ``best'' DF;
\item and then take this DF to be representative of {\it all} of the
  DFs for the assumed potential.
\end{enumerate}
The first step is fine (at least for certain applications), but the
second is wholly unjustified and ignores the fact that $\chi^2[\b f]$
is hugely degenerate.

To see a variant of this problem in a much milder context, consider
how one measures $M_\bullet$ in real galaxies.  The potential then has
at least one additional free parameter, the mass-to-light
ratio~$\Upsilon$, and one has to construct a grid of models for a
range of different values of $M_\bullet$ and $\Upsilon$.  The
uncertainties on $M_\bullet$ are never obtained by picking out a
special value of $\Upsilon$ for each $M_\bullet$; instead one
marginalizes $\Upsilon$ either explicitly \citep[e.g.,][]{g03} or
implicitly through the use of a $\Delta\chi^2$ criterion
\citep[e.g.,][]{vdm98,capp02}.  This idea of marginalization is key to
resolving the issues noted above.

\section{Fitting models to observations: a Bayesian approach}

The maximum-likelihood approach of the previous section pays scant
attention to the orbit weights~$f_i$, which serve merely as
co-ordinates used to locate a point somewhere along the degenerate
minimum in $\chi^2$.
From a
Bayesian point of view, however, the $f_i$ are {\it nuisance
  parameters}.  Although we are not interested knowing their precise
values, they deserve to be treated on an equal footing with the
parameters defining the potential.

Applying Bayes' theorem twice, the
posterior probability of a model with potential $\psi$ and a set of
orbit weights~$\b f$,
\begin{equation}
  \label{eq:bayesone}
  \pr(\psi,\b f|D) \propto \pr(D|\b f,\psi) \pr(\b f|\psi)\pr(\psi),
\end{equation}
where $\pr(D|\psi,f)=\exp[-{1\over2}\chi^2(\b f|\psi)]$ is the usual likelihood
and the priors $\pr(\b f|\psi)$ and $\pr(\psi)$ will be discussed
later.
Since we are not interested in the values of weights (as long as they
are non-negative), let us marginalize~(\ref{eq:bayesone}) to obtain
\begin{equation}
  \label{eq:bayestwo}
  \pr(\psi|D) \propto \pr(D|\psi) \pr(\psi),
\end{equation}
where the marginalized likelihood,
\begin{equation}
\begin{split}
  \label{eq:marglike}
  \pr(D|\psi) &\equiv \int \pr(D|\b f,\psi) \pr(\b f|\psi)\,\d\b f\\
&\equiv \exp\left[-{1\over2}\chi^2_{\rm marg}\right],
\end{split}
\end{equation}
is obtained by summing the likelihood over all non-negative DFs, each
weighted by the as-yet-unspecified prior $\pr(\b f|\psi)$.  Notice
that this is directly analogous to the partition function in
statistical mechanics, with $\chi^2$ playing the role of energy and
the prior standing in for the density of states.
The conventional maximum-likelihood
method of the last section can be viewed as the very crude
approximation
\begin{equation}
\pr(D|\psi) \sim \max_{\{\b f\}\ge0}\pr(D|\b f,\psi)=
\exp\left[-{1\over2}\min_{\b f\ge0}\chi^2[\b f]\right],
\end{equation}
obtained by completely ignoring the prior $\pr(f|\psi)$ and
approximating the remaining integral by the peak value of its
integrand.  There is a straightforward and obvious analogue for the
maximum penalized likelihood method.

\subsection{The priors}
\label{sec:priors}

There is nothing noteworthy about the choice of the potential prior
$\pr(\psi)$ for the situation considered here.  The potential has one
free parameter, $M_\bullet$, which can be zero or positive, meaning
that the natural prior to use is flat in $\log M_\bullet$.

The choice of prior for the DF, $\pr(\b f|\psi)$, is more interesting.
Recall that we use discrete cells~(\ref{eq:DFdiscretization}) to model
continuous phase space.  Now, there is no a priori natural way to
partition phase space into cells.  Let us assume for the moment that
there are no correlations among cells and that the prior is independent
of location in phase space.  Let $\mu_i$ be the prior expectation
value for the dimensionless luminosity $F_i\equiv L_i/L_{\rm s}$
(eq.~\ref{eq:flux}) in cell~$i$; it can therefore be thought of as measure
of the cell's volume.
Then a natural requirement on
the prior is that, given a partition~$\pi$ of phase space, we should
be able to select any cell and construct a new partition~$\pi'$ by
subdividing this cell into $m>0$ subcells and have that
\begin{equation}
\begin{split}
\label{eq:id}
\pr_\pi(F|\mu,\psi) &=
\int_0^F\!\! \d F_1\cdots\int_0^F\!\!\d F_m \,\delta(F_1+\cdots +F_m-F)\\
& \qquad\times\pr_{\pi'}(F_1|\mu_1,\psi)\times\cdots\times
\pr_{\pi'}(F_m|\mu_m,\psi),
\end{split}
\end{equation}
where the cell volumes satisfy $\mu_1+\cdots+\mu_m=\mu$.  That is,
marginalizing over the subcells should return the original prior.  A
consequence of this is that, for finite-resolution data, the
marginalized likelihood $\pr(D|\psi)$ is independent of the chosen
partition provided only that one uses fine enough cells.

The infinite-divisibility (hereafter ID) condition~(\ref{eq:id}) puts
strong constraints on the form of the prior.  It is not
satisfied by any of the simplest commonly used priors
\begin{equation}
\pr(F|\psi)\propto
\begin{cases}
1, & \text{(uniform)}\\
1/F,        & \text{(Jeffreys),}\\
\exp[-\alpha F\ln F] & \text{(entropy).}
\end{cases}
\end{equation}
The form of the convolution in eq.~(\ref{eq:id}) suggests that Laplace
transforms might be helpful in finding ID priors, and indeed it can be
shown \citep[][\S XIII.7]{feller} that a probability distribution $\pr(F|\mu)$
satisfies~(\ref{eq:id}) if and only if its Laplace transform
\begin{equation}
  \tilde\pr(s|\mu)\equiv \int_0^\infty\d F\, \e^{-s F} \pr(F|\mu)
\end{equation}
is of the form
\begin{equation}
  \label{eq:levykhinchin}
  \tilde\pr(s|\mu) = \exp\left[-\int_0^\infty {1-\e^{-s F}\over F}
    {\cal M}(\mu,\d F)\right],
\end{equation}
with the only constraint on the measure ${\cal M}(\mu,\d F)$, known as
the {\it L\'evy measure}, being that the integral
\begin{equation}
  \int_\epsilon^\infty F^{-1}{\cal M}(\mu,\d F)
\end{equation}
converges for all $\epsilon>0$.    Thus we can construct priors that are
guaranteed to be ID simply by considering a variety of choices for
${\cal M}$.  For example, substituting ${\cal M}(\mu,\d F) =
\mu\delta(F-1)\,\d F$ in~(\ref{eq:levykhinchin}) results in the Possion
distribution, while ${\cal M}(\mu,\d F) =\mu\e^{-F}\,\d F$ gives the gamma
distribution.  Both of these obviously satisfy the
ID criterion~(\ref{eq:id}).

A concise and very readable introduction to this subject is given by
\citet{skilling98}.  He argues that the maximally ignorant choice of
${\cal M}$ when one knows only a characteristic scale for~$F$ is
\begin{equation}
{\cal M}(\mu,\d F) = \mu F\e^{-F}\,\d F.
\end{equation}
This results in the so-called ``massive inference'' prior 
\begin{equation}
\begin{split}
\pr(F_i|\mu_i,\psi) &= \e^{-\mu_i}
  \Bigg[ \delta\left(F_i\right) \\
&\qquad + \exp\left(-{F_i}\right)\sqrt{\mu_i\over F_i}\,
    I_1\left(2\sqrt{\mu_iF_i}\right)
  \Bigg],
\end{split}
\label{eq:incontinent}
\end{equation}
where $I_1$ is the first-order modified Bessel function of the first
kind.  Appendix~B gives an elementary derivation of this prior,
explaining how it is the natural generalization of entropy to
continuous distributions.  I adopt it for the calculations below.
\begin{figure}
  \begin{center}\includegraphics[width=0.8\hsize]{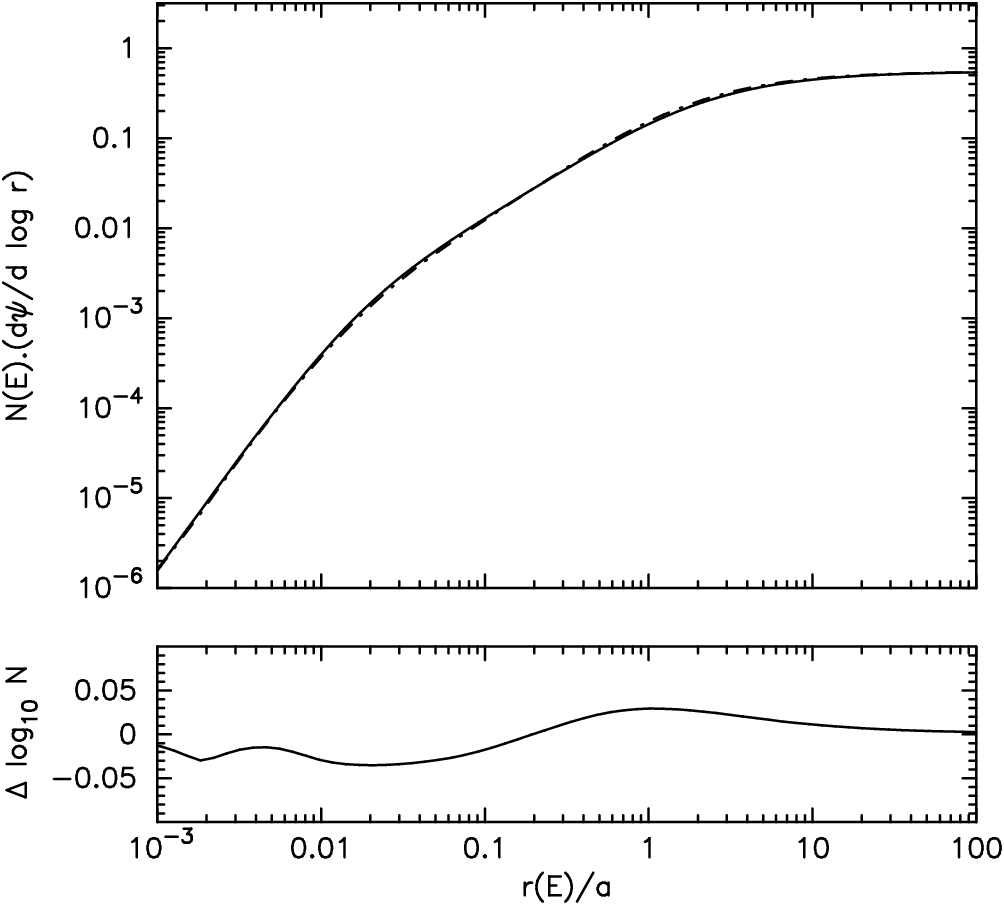}\end{center}
  \caption{The differential energy distribution $N(\E)$
    (equ.~\ref{eq:en}) for isotropic (solid curve) and anisotropic
    (dashed) toy galaxies, along with their difference (bottom panel).
    $r(\E)$ is the apocentre radius of a perfectly radial orbit with
    energy~$\E$.}
  \label{fig:ne}
\end{figure}

There remains the question of what to choose for the prior
weights~$\mu_i$.  From the Laplace
transform~(\ref{eq:levykhinchin}) it is straightforward to show that
the prior mean and variance,
\begin{equation}
 \label{eq:priormeanvar}
  \begin{split}
E(F) &= \int_0^\infty {\cal M}(\mu,\d F), \\    
{\rm var}(F) & = \int_0^\infty F{\cal M}(\mu,\d F),
  \end{split}
\end{equation}
and therefore that the prior~(\ref{eq:incontinent}) likes to have
$F\approx \mu\pm\sqrt{2\mu}$.  In the absence of a compelling model of
galaxy formation, I simply use the well-known fact \citep{bt} that for
reasonable galaxy models the differential energy distribution,
\begin{equation}
  \label{eq:en}
N(\E) =\int_0^{J^2_{\rm max}(\E)} f(\E,J^2) g(\E,J^2)\d J^2,
\end{equation}
is almost independent of anisotropy once the galaxy's luminosity
density $j(r)$ and potential~$\psi(r)$ have been specified (e.g., see
fig.~\ref{fig:ne}).  For each trial potential I find the
isotropic DF $f_{\rm iso}(\E)$ that produces the luminosity
density~(\ref{eq:lumprof}) and, following~(\ref{eq:lumcell}) and~(\ref{eq:flux}), assign
\begin{equation}
  \mu_i = {1\over L_{\rm s}}\cdot \int_{V_i} f_{\rm iso}(\E)g(\E,J^2)\,\d\E\d J^2.
\label{eq:mu}
\end{equation}
This depends on the choice of characteristic luminosity scale $L_{\rm
  s}$ (equ.~\ref{eq:flux}).  Now, the prior RMS fractional spread in each
cell is equal to $\sqrt{2/\mu_i}$, which clearly depends on $L_{\rm s}$
and has a Poisson-like $N^{-1/2}$ dependence on the cell
volume.\footnote{This is inevitable for any ID distribution, which can
  be seen either from~(\ref{eq:priormeanvar}) or 
  from the fact that any ID prior is a limit of a sequence of compound
  Poisson distributions \citep{feller}.}  In order to make the prior
variance independent of the partitioning scheme used to represent the
DF, I introduce a second, independent reference partition and choose
\begin{equation}
  \label{eq:Lscale}
  L_{\rm s}(\E,J^2) = {2\over\delta^2} \int_{V_{\rm ref}(\E,J^2)} f_{\rm iso}(\E)g(\E,J^2)\,\d\E\d J^2,
\end{equation}
so that the prior RMS fractional spread in each reference cell is
given by the adjustable parameter~$\delta$.  For the results presented
here I use the partition defined by the $n_{\E}\times n_J=100\times10$
grid for this reference partition.

\subsection{Marginalization}
\label{sec:marginalization}

Although one could attempt the difficult task of evaluating the
marginalized likelihood $\pr(D|\psi)$~(eq.~\ref{eq:marglike}) directly, we
are not so much interested in the absolute value of $\pr(D|\psi)$ as in
the odds,
\begin{equation}
  \label{eq:odds}
  {\pr(\psi_1|D)\over\pr(\psi_0|D)} = 
{\pr(D|\psi_1)\pr(\psi_1)\over\pr(D|\psi_0)\pr(\psi_0)},
\end{equation}
of one potential~$\psi_1$ compared to another~$\psi_0$.  I evaluate
the ratio $\pr(D|\psi_1)/\pr(D|\psi_0)$ using the method of
thermodynamic integration \citep{neal93}.  Consider two models, one
having potential~$\psi_0$, the other with a slightly different
potential~$\psi_1$, but both having $n_\E\times n_J$ DF components
chosen according to the scheme described in section~3 for their
respective potentials.  Let
\begin{equation}
  \label{eq:intermed}
  Z_\lambda \equiv \int \exp [C_\lambda(\b f)]\,\d\b f
\end{equation}
where 
\begin{equation}
  \label{eq:intermedntg}
\begin{split}
  C_\lambda(\b f) &\equiv
 -{1\over2}\left[(1-\lambda)\chi^2(\b f|\psi_0) + \lambda\chi^2(\b f|\psi_1)
 \right]\\
& \quad+ (1-\lambda)\ln\pr(\b f|\psi_0) + \lambda \ln\pr(\b f|\psi_1).
\end{split}
\end{equation}
As the parameter $\lambda$ varies between 0 and~1, $Z_\lambda$
interpolates smoothly between $\pr(D|\psi_0)$ and $\pr(D|\psi_1)$.
Taking the logarithm of~(\ref{eq:intermed}) and differentiating with
respect to~$\lambda$,
\begin{equation}
\begin{split}
  \label{eq:zprime}
  {\d\over\d\lambda}\log Z_\lambda &= \int {\d C_\lambda\over
    \d\lambda}\cdot {1\over Z_\lambda}\exp[C_\lambda(\b f)]\d\b f\\
& = \big\langle {\d C_\lambda\over \d\lambda}\big\rangle_\lambda,
\end{split}
\end{equation}
where $\langle C\rangle_\lambda$ denotes the expectation value of
$C(\b f)$ when $\b f$ has probability density $\exp[C_\lambda(\b
f)]/Z_\lambda$.  Therefore we can use a Markov-Chain Monte Carlo
method to draw points from this density, and taking the mean value of
$\d C_\lambda/\d\lambda$ for these points gives an immediate estimate
of $\d\log Z_\lambda/\d\lambda$.  Then, integrating,
\begin{equation}
  \begin{split}
 \int_0^1\d\lambda\cdot \left({\d\over\d\lambda}\log Z_\lambda\right)
  &= \log\left(Z_1\over Z_0\right)
  = \log\left(\pr(D|\psi_1)\over\pr(D|\psi_0)\right)\\
  &= -{1\over2}\left(\chi^2_{\rm marg}(\psi_1)-\chi^2_{\rm
       marg}(\psi_0)\right).
  \end{split}
\label{eq:margdiff}
\end{equation}
This method will be reasonably efficient only if the dependence of
$C_\lambda(\b f)$ on $\b f$ does not vary significantly as $\lambda$
changes, which is the case for the choice of $f_i$ using the scheme
described in Section~3.

For the results presented below I use a Gibbs
sampler to draw $5\times10^6$ points from $\exp[C_\lambda(\b f)]$
after a burn-in period of $10^6$ iterations
starting from $\b F=\b\mu$.  
 Instead of evaluating the
derivative~(\ref{eq:zprime}) for a range of fixed $\lambda$ I instead
increase $\lambda$ slowly from 0 to 1 over the course the iterations,
yielding $\log \pr(D|\psi_1)-\log\pr(D|\psi_0)$ directly.  A direct
test of this procedure is to run it backwards by swapping the
potentials around and calculating $\log
\pr(D|\psi_0)-\log\pr(D|\psi_1)$.  I find that that both forward and
backward iterations typically agree very well, provided
$\delta\la20$.  For larger $\delta$, the delta function in the
prior~(\ref{eq:incontinent}) becomes dominant and the posterior
becomes effectively stuck at $F_i=0$ for a significant fraction of the
DF components.

\subsection{Results}

\begin{figure*}
\begin{center}
\includegraphics[width=0.4\hsize]{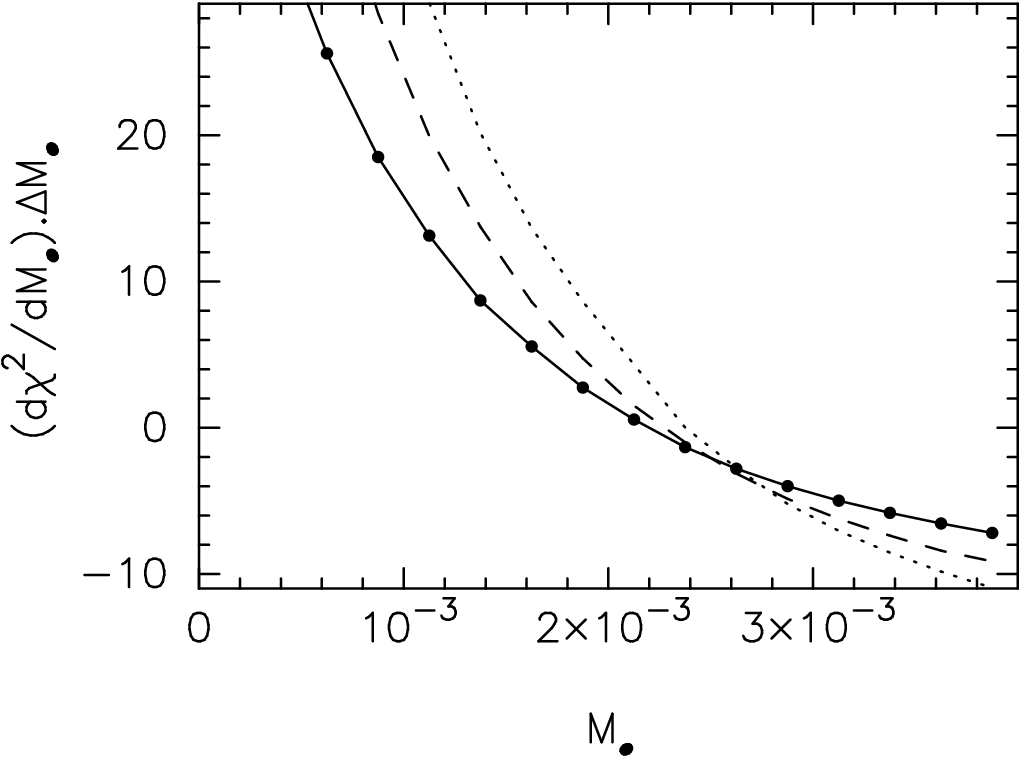}
\includegraphics[width=0.4\hsize]{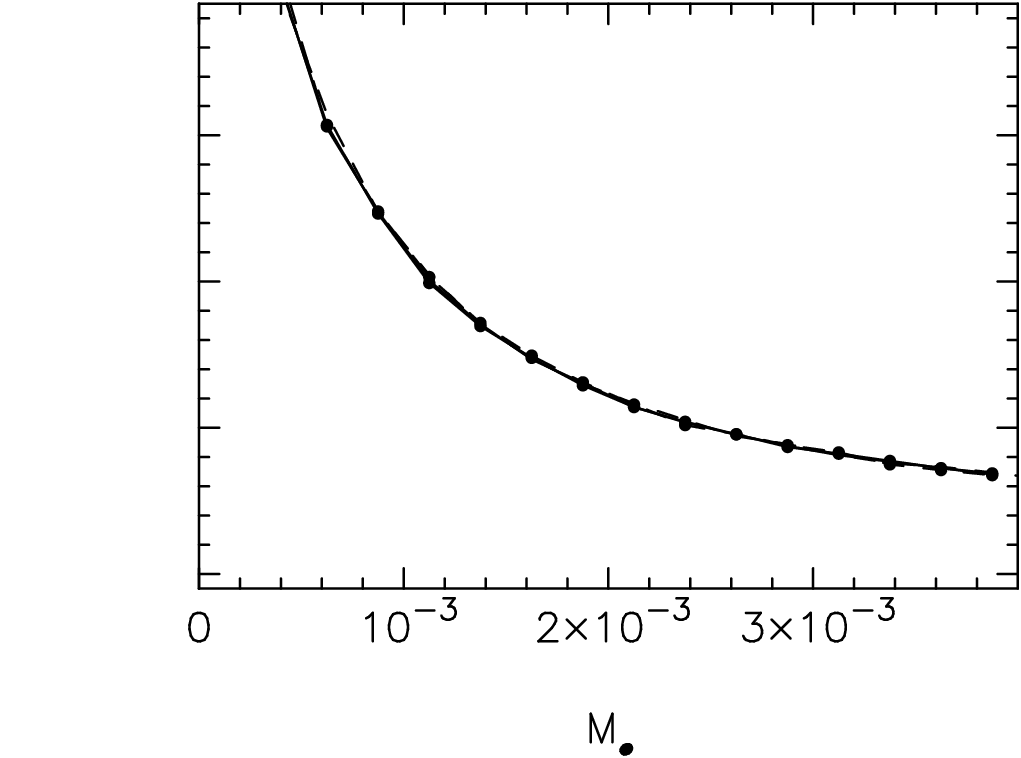}
\end{center}
\caption{The importance of infinite divisibility.  The left panel
  plots the derivatives of the marginalized log-likelihood
  $\chi^2_{\rm marg}(M_\bullet)=-2\ln\pr(D|M_\bullet)$ calculated
  using 
  equ.~(\ref{eq:margdiff}) for a non-ID flat prior.  Solid, dashed and
  dotted curves correspond to models with $n_{\E}\times
  n_J=100\times10$, $200\times10$ and $200\times20$ DF components,
  respectively.  The panel on the right plots the same for the ID
  prior~(\ref{eq:incontinent}) with $\delta=5$.  }
\label{fig:compareflat}
\end{figure*}

\begin{figure}
\begin{center}
\includegraphics[width=0.8\hsize]{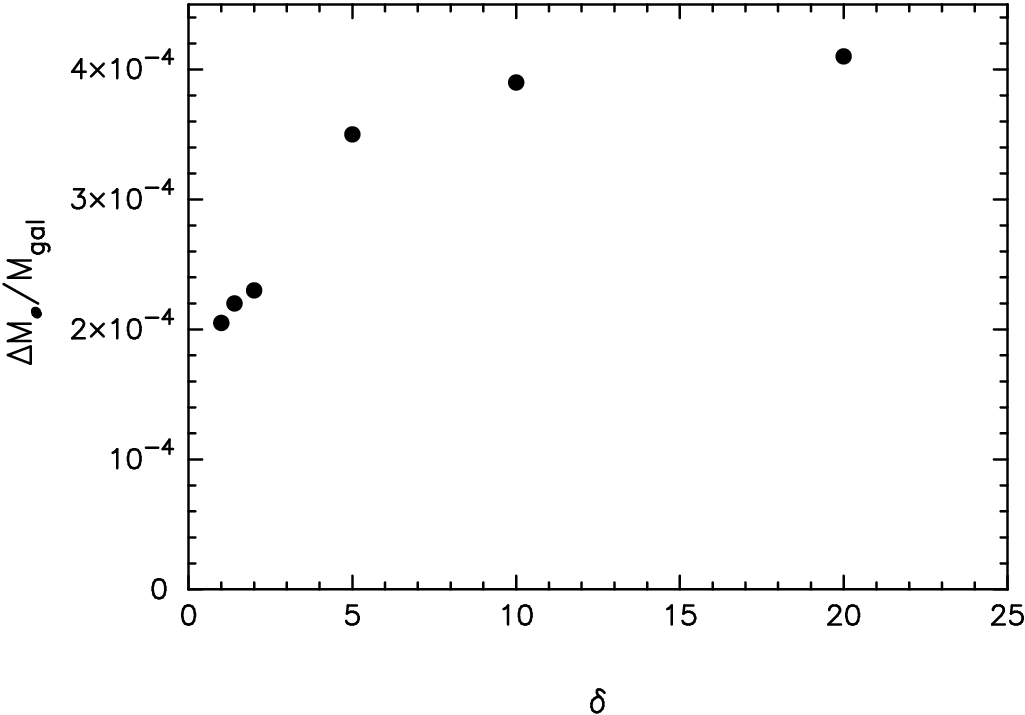}
\end{center}
\caption{The dependence of the formal uncertainty in BH mass on the
  fractional variance $\delta^2$ used in the prior.}
\label{fig:deltadep}
\end{figure}

\begin{figure}
\begin{center}
\includegraphics[width=0.8\hsize]{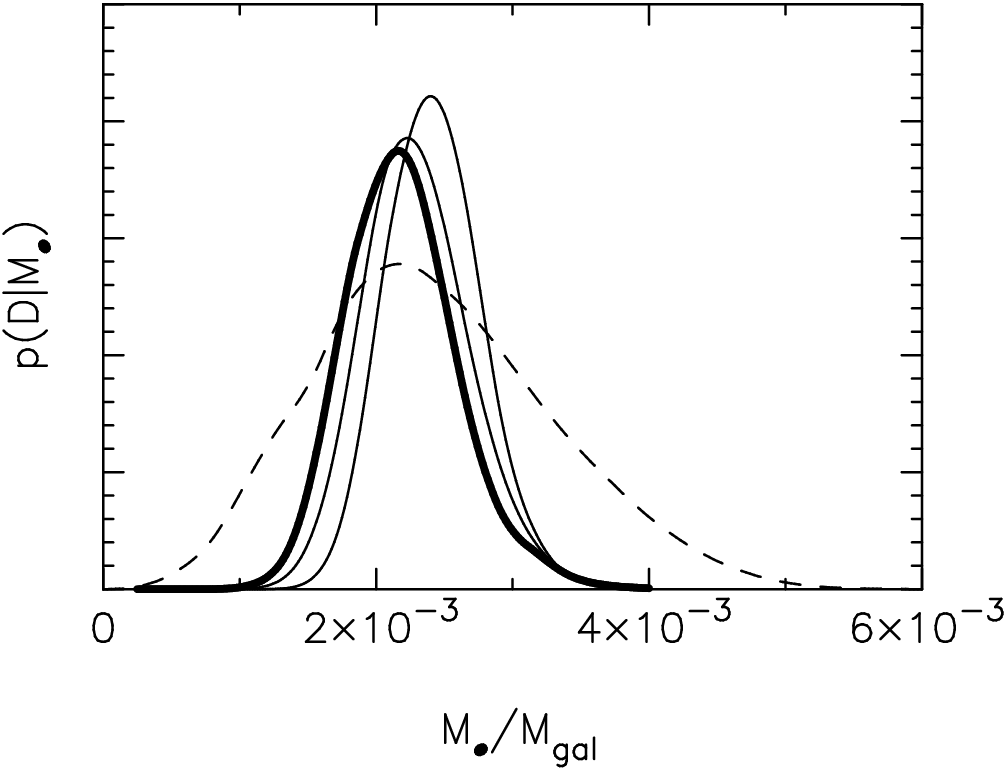}
\end{center}
\caption{Marginalized likelihoods for one realization
  of the anisotropic toy galaxy.  The solid curves plot
  $\pr(D|M_\bullet)$ using priors with $\delta=5$, 10 and (heavy solid
  curve) 20.  For comparison, the dashed curve plots the corresponding
  probability distribution returned by the conventional
  maximum-likelihood method.}
\label{fig:margprob}
\end{figure}

Here I present results of applying the Bayesian analysis to
observations of the anisotropic toy galaxy.  The results for the
isotropic galaxy are similar but less instructive since I use the
isotropic case to assign the prior weights $\b\mu$.

Figure~\ref{fig:compareflat} demonstrates the importance of using a
prior that satisfies the ID criterion~(\ref{eq:id}).  Taking a
(non-ID) prior flat in the orbit weights~$\b f$ leads to a
marginalized likelihood~$\pr(D|\psi)$ that depends on the number of
orbits used and, more generally, on exactly how one discretizes phase
space.  In contrast, the marginalized likelihood $\pr(D|M_\bullet)$ for the
ID prior~(\ref{eq:incontinent}) does not depend on whether we
discretize using $n_\E\times n_J=100\times10$, $200\times10$ or
$200\times20$ cells.  Incidentally, this also shows that, for the
present purposes at least, it is acceptable to use delta
functions~(eq.~\ref{eq:DFdiscretization}) to calculate the contribution
each cell makes to the observations.

Although the marginalized likelihood is independent of the
discretization, it still has one free parameter, the fractional
variance per reference cell~$\delta^2$.  To investigate the dependence
of the results on $\delta$, I use the mean and variance of the
posterior distribution $\pr(M_\bullet|D,\delta)$,
\begin{equation}
  \begin{split}
  \overline M_\bullet(\delta) &= \int M_\bullet
    \pr(D|M_\bullet,\delta)\pr(M_\bullet)\,\d M_\bullet\\
  (\Delta M_\bullet(\delta))^2 & = \int
      \left(M_\bullet-{\overline M_\bullet}(\delta)\right)^2
       \pr(D|M_\bullet,\delta)\pr(M_\bullet)\,\d M_\bullet.
  \end{split}
  \label{eq:bayesmom}
\end{equation}
Figure~\ref{fig:deltadep} shows how the formal uncertainty $\Delta
M_\bullet$ varies with $\delta$ for a typical realization of the
observations.  As one might expect, $\Delta M_\bullet$ is low for
small $\delta\sim1$, but grows rapidly as $\delta$ increases.  Once
$\delta$ reaches $\sim10$ though, $\Delta M_\bullet$ stabilizes at
about $0.4\times10^{-3}\,M_{\rm gal}$.  The marginalized likelihoods
$\pr(D|M_\bullet)$ used in calculating $\Delta M_\bullet$ for the
three largest values of $\delta$ plotted are shown in
Figure~\ref{fig:margprob}, which shows that there is little change as
one increases $\delta$ from 10 to 20.  It is somewhat surprising that
$\delta$ needs to be so large.  This might be a consequence of the
power-law dependence of the DF (eq.~\ref{eq:DF}) on~$J$.  For
comparison, the figure also includes the probability distribution
returned by applying the conventional maximum-likelihood method to the
same dataset.  Taking many realizations of the galaxy, the
Bayesian method yields a mean $\Delta M_\bullet$ of
$0.44\times10^{-3}M_{\rm gal}$ compared to the
$0.99\times10^{-3}M_{\rm gal}$ returned by the conventional method (table~1).

\section{Conclusions}

Most BH mass estimates come from the conventional maximum-likelihood
method.  They fit observations using models with specially constructed,
unrealistically spiky DFs, yielding implausibly good fits to the
observations.  Results from the toy galaxies considered here suggest
that the (unpenalized) maximum-likelihood method nevertheless does
yield reliable BH masses, but with overly pessimistic error estimates.

I confirm VME04's result that a wide range of BH masses can yield
perfect fits to finite-resolution, noiseless observations.  Contrary
to their somewhat speculative arguments, however, I show that one does
not expect to find flat-bottomed $\chi^2$ distributions in practice,
unless one is blessed with very high signal-to-noise observations
(figure~\ref{fig:varynoise}) and fits only the low-order shapes of the
VPs.

Although the maximum-likelihood method yields reliable BH masses, it
is flawed because it considers only one DF for each BH mass.  The
remedy is to consider all possible DFs for each potential, weighting
each one by a suitably chosen prior.  This significantly improves the
constraints on the BH mass, since the closer a trial potential comes
to the true potential, the greater the number of non-negative DFs that
are consistent with the observations.

\subsection{Open questions and future work}

\subsubsection{Choice of prior}
An open question is to what extent these results depend on the choice
of prior weights (equ.~\ref{eq:mu}), and more generally on the use of
the prior~(\ref{eq:incontinent}), which is just one of many possible
infinitely divisible distributions.  A very general argument
\citep{kingman} shows that drawing random realizations from most ID
priors will yield spiky, uncorrelated distributions.  This is
just what one expects if dealing with galaxies at the level of
individual stars, but the present models are  far from this level
of detail and one might plausibly expect some degree of correlation
among neighbouring cells in phase space.  Making this idea
quantitative is difficult, however.

\subsubsection{ Computational scheme}
It would be straightforward in principle to apply the ideas presented
here to axisymmetric or triaxial galaxy models.  One could simply
adopt the prior~(\ref{eq:incontinent}), using the scheme described by
\citet{jens} to calculate the volumes used to assign the prior weights
$\mu$.  In practice, however, calculating the marginalized likelihood
$\pr(D|\psi)$ will probably be very difficult.  The DFs of
axisymmetric galaxies are three-integral, which means that, in order
to ensure a fine enough discretization, one has to use many more DF
components than for two-integral spherical models, with the Markov
Chain Monte Carlo procedure used in section~\ref{sec:marginalization}
taking correspondingly more iterations to converge.  On the other
hand, it is very likely that there are much more efficient methods
than the combination of Gibbs sampling and thermodynamic integration
used here.

\subsubsection{More immediate problems}
Before applying this method to real galaxies though, it is probably
worth addressing the following more tractable problems first:
\begin{enumerate}
\item VPs are extracted from spectra, which suffer from poorly
  understood systematic errors \citep{ryan06}.
\item For real observations, neither individual VP velocity bins nor
  (surprisingly) Gauss--Hermite coefficients are independent
  \citep{ryan06}.
\item Most models of axisymmetric galaxies make an ad hoc assumption
  about the galaxy's three-dimensional light distribution $j(R,z)$,
  despite the fact that there are many $j(R,z)$ consistent with a
  given surface brightness distribution \citep[e.g.,][]{kochanekrybicki}.
  Neglect of this degeneracy can lead to incorrect inferences about
  the galaxy's orbit structure \citep{depropaper} which are likely to
  affect BH mass estimates.
\item The widely used
  $\Delta\chi^2$ criteria for obtaining
  uncertainties on BH masses are based on the assumption that
  $\chi^2(M_\bullet,\Upsilon)$ is close to quadratic, which does not
  necessarily hold in practice.
\end{enumerate}

\section*{Acknowledgments}

I thank James Binney, Karl Gebhardt, Andrew Jaffe, Douglas Richstone,
Prasenjit Saha and Scott Tremaine for helpful discussions, the anonymous
referee and the participants of the 2006 Lorentz Center workshop on
Galactic Nuclei for comments that greatly improved the presentation of
the results contained here, and the Royal Society for financial support.

\appendix

\section[]{Observables of DF components}

I use a rectangular $(x,y,z)$ co-ordinate system with origin~$O$ at
the galaxy centre and whose $Oz$ axis is parallel to lines of sight.
Points on the plane of the sky are then labelled by the co-ordinates
$(x,y)$.  Of course, real observations do not have perfect spatial
resolution.  Instead, any function $f(x,y)$ defined on the plane of
the sky is measured convolved with a two-dimensional point-spread
function $\psf(\Delta x,\Delta y)$:
\begin{equation}
f_{\rm meas}(x,y) = \int\int\psf(x-x',y-y')f(x',y')\,\d x'\d y'.
\end{equation}
Since we assume that the galaxy is spherical, then $f(x,y)=f(R)$,
where $R=\sqrt{x^2+y^2}$ is the usual cylindrical polar radius,
and the psf-convolved value of $f$ at radius~$R$,
\begin{equation}
f_{\rm meas}(R) = \int p(R,R')f(R')\,R'\d R',
\label{eq:psfconvolve}
\end{equation}
where 
\begin{equation}
p(R,R')\equiv \int \psf(R-R'\cos\phi',R'\sin\phi')\,\d\phi'
\label{eq:psfRR}
\end{equation}
is the azimuthally integrated contribution of light from radii~$R'$ to
measurements at radius~$R$.  
For example, in section~\ref{sec:toy} the toy galaxies are ``observed''
through annulii that admit light between some radii
$R_1$ and~$R_2$.  For this situation
\begin{equation}
p(R,R') = {2\over|R_2^2-R_1^2|} \times
\begin{cases}
 1 & \text{if $R_1<R<R_2$,}\\
 0 & \text{otherwise.}
\end{cases}
\end{equation}
A more realistic psf is a Gaussian with some
dispersion~$\sigma_\star$, for which
\begin{equation}
 p(R,R') = {1\over\sigma_\star^2}\exp\left[-{{R^2+R'^2}\over2\sigma_\star^2}\right]
 I_0\left(RR'\over\sigma_\star^2\right),
\end{equation}
where $I_0$ is a Bessel function.  Both these examples are symmetric,
but we note that we can use (\ref{eq:psfconvolve})
and~(\ref{eq:psfRR}) to convolve any spherically symmetric function
$f(R)$ with a psf of arbitrary shape.

Now consider a single DF component~(\ref{eq:DFdiscretization}) of
energy~$\E$ and angular momentum~$J$ per unit mass in
potential~$\psi(r)$.  Written explicitly as a function of $(\b x, \b
v)$,
\begin{equation}
f(\b x, \b v) =
\delta\big[\psi-{1\over2}(v_r^2+v_\theta^2+v_\phi^2)-\E\big]
\delta\big[r^2(v_\theta^2+v_\phi^2)-J^2\big].
\label{eq:oneDF}
\end{equation}
The individual orbits making up the DF component have
peri- and apo-centre radii $r_\pm$ given by the
roots of the equation $V_r(r)=0$, where
\begin{equation}
V_r^2(r)\equiv 2[\psi(r)-\E]-{L^2\over r^2},
\end{equation}
and I have omitted the obvious dependence of the result on $\E$,
$J$ and~$\psi$.  The velocity moments of the component~(\ref{eq:oneDF}),
\begin{align}
  [v_r^{2i} v_\theta^{2j}v_\phi^{2k}](r) & \equiv
  \int\d^3\b v\, v_r^{2i} v_\theta^{2j}v_\phi^{2k}\, f\cr
  &= 2B\left(\textstyle{i+{1\over2},j+{1\over2}}\right)\cr
 &\quad
\times
\begin{cases}
  {V_r^{2i-1}J^{2(j+k)}\over
    r^{2(j+k+1)}} & \text{if $r_-<r<r_+$,}\\
    0 & \text{otherwise.}
\end{cases}
\end{align}
Taking $i=j=k=0$ yields the luminosity density $j(r)=2\pi/r^2V_r$,
which has integrable singularities at both $r=r_-$ and $r=r_+$.
Integrating $j(r)$ over radius, the total luminosity of the
component~(\ref{eq:oneDF}) is given by
\begin{equation}
L = 8\pi^2\int_{r_-}^{r_+} {\d r\over V_r},
\label{eq:dosfac}
\end{equation}
which is just the usual density-of-states factor.

Substituting~$j(r)$ into equation~(\ref{eq:psfconvolve}), the
psf-convolved surface brightness distribution
\begin{equation}
\begin{split}
I(R) &= 4\pi\int_0^\infty \d R'\,p(R,R')R' \int_{-\infty}^\infty {\d z'\over
  (R'^2+z'^2)V_r(R',z')}\\
&= 4\pi\int_{r_-}^{r_+} {\d r\over
  r^2V_r}\int_0^{2\pi}p(R,r\sin\theta)\sin\theta\,\d\theta.
\end{split}
\end{equation}
I evaluate this integral numerically by substituting
$r=r_-+(r_+-r_-)\sin^2u$ and applying Simpson's rule with intervals
$\Delta u=\pi/200$ and $\Delta\theta=\pi/50$.  A simple test of
this calculation is to compare the integrated surface brightness
$2\pi\int R I(R)\,\d R$ against the density-of-states
factor~(\ref{eq:dosfac}).  I find that the two typically agree to
around one part in $10^4$.

The calculation of velocity profiles is a little more involved.
A star at position $(R,z)$ with velocity $(v_r,v_\theta,v_\phi)$
has projected line-of-sight velocity
\begin{equation}
\vp = {1\over r}\left[zv_r+Rv_\theta\right].
\end{equation}
The luminosity density of stars located at a position $(R,z)$ having
projected velocities in the range $v_1<\vp<v_2$ is
\begin{equation}
\begin{split}
j(R,z;v_1,v_2)\!\!\! & = 
\int_{v_1}^{v_2}\d\vp\int f(\b x,\b v)
\delta\left[{1\over r}\left(zv_r+Rv_\theta\right)-\vp\right]\,\d^3\b v\\
& = \sum_\pm\int{\d\vp\over Rr|V_rV_{\phi\pm}|},
\label{eq:histbinRz}
\end{split}
\end{equation}
where the range of integration includes only those $\vp\in[v_1,v_2]$
for which
\begin{equation}
V_{\phi\pm}^2(R,z;\vp)\equiv{J^2\over r^2}-{(r\vp\pm zV_r)^2\over R^2}
\label{eq:vphipm}
\end{equation}
is non-negative, and the $\sum_\pm$ takes care of both possibilities
for the sign of $v_r$.  Integrating along the line of sight and
using~(\ref{eq:psfconvolve}), the (unnormalized) psf-convolved
contribution of the DF component~(\ref{eq:oneDF}) to a VP bin that
extends from from $\vp=v_1$ to~$v_2$ is
\begin{equation}
\begin{split}
\L(R;v_1,v_2) &= 2\int_{r_-}^{r_+} r\d r\times \\
&\quad
\int_0^{2\pi} p(R,r\sin\theta)
j(R,r\cos\theta;v_1,v_2)\d\theta.
\label{eq:histbinR}
\end{split}
\end{equation}
The integral~(\ref{eq:histbinRz}) for $j(R,r\cos\theta;v_1,v_2)$ can
be carried out by hand, but finding efficiently the regions where it
is nonzero is far from easy.  So, to evaluate the double
integral~(\ref{eq:histbinR}) I simply adopt the scheme I use to
calculate~$I(R)$ and substitute $r=r_-+(r_+-r_-)\sin^2u$: the
integration over $\vp$ already takes care of the worst effects of the
singularities in the integrand of~(\ref{eq:histbinRz}).  To test this
approach, I compare the integrated VP histogram, $\sum_i\L(R;i\Delta
v,(i+1)\Delta v)$ for some choice of~$\Delta v$, against~$I(R)$.  For
my standard integration parameters, the typical RMS fractional
difference between these two quantities is about $10^{-6}$.  This
merely indicates that very little light ``leaks'' from the models, not
that its VPs are calculated to that accuracy.  Comparing VPs
calculated with different stepsizes, I estimate that this standard
integration scheme yields VPs with an RMS fractional error of a few
parts in $10^4$.

To obtain the contribution each DF component makes to the modifed
moments $Ih_i(R)$, 
I calculate the component's VP histogram
$\L(R;v_i,v_{i+1})$ for 20 bins from $v_0=0$ to $v_{24}=5\sigma$,
where $\sigma$ is the velocity dispersion used in the Gauss--Hermite
expansion, and use this histogram in equation~(\ref{eq:GHcoeff}).


\section{Priors}
The following derivation of the prior~(\ref{eq:incontinent}) follows
\citet{skilling98}.  Imagine a monkey \citep{gulldaniell} with a bag
of $N$ identical stars each of luminosity~$L_\star$.  He sits outside
a big box containing phase space (strictly, integral space), takes
each star in turn and throws it into the box.  The probability of each
star landing in a small volume $\d^3\b x\d^3\b v$ around the point
$(\b x,\b v)$ is $\alpha(\b x,\b v)\d^3\b x\d^3\b v$.  If we take a
small cell of volume $\delta V$, the probability that $r$ of the $N$
stars land inside the cell is
\begin{equation}
\label{eq:binom}
\pr(r|N,\alpha\delta V) = {N\choose r}(\mu\delta V)^r(1-\alpha\delta V)^{N-r}.
\end{equation}
Now shrink the cell volume $\delta V\to0$ and increase the number of stars
$N\to\infty$ keeping the product $N\delta V$ constant.
Equation~(\ref{eq:binom}) becomes
\begin{equation}
\pr(r|\alpha) = {\alpha^r\over
  r!}\e^{-\mu},
\label{eq:poisson}
\end{equation}
describing a Poisson process with mean $\mu\equiv N\alpha\delta V$.
Therefore, the probability of having luminosity~$L_i$ in a cell of
phase-space volume $V_i$ is given by
\begin{equation}
\label{eq:monkeyfac}
\pr(L_i|\mu_i) = \e^{-\mu_i} \sum_{r=0}^\infty {\mu_i^r\over
  r!}\delta(L_i-r L_\star),
\end{equation}
with $\mu_i\equiv N\int_{V_i}\alpha \d^3\b x\d^3\b v\simeq N\alpha_i V_i$.
For a partition of phase space into $n$ cells, the probability of the
configuration $(L_1,\ldots,L_n)$ is given by a product of factors
like~(\ref{eq:monkeyfac}),
which obviously satisfies the criterion~(\ref{eq:id}) for infinite
divisibility.  Apart from uninteresting scale factors, this prior is
identical to equation~(20) of \citet{Shu78}.  In the limit $\mu_i\gg1$
it takes on the entropy-like form
$\e^{-\mu}\exp[-(L/L_\star)(1+\ln(L/\alpha L_\star))]$.

The prior~(\ref{eq:monkeyfac}) has the disadvantage of requiring a
discretization of the light into individual stars, with the result
that $L$ is very strongly peaked at 0 whenever $\mu\ll1$.  For an
alternative, suppose that we place the monkey inside the box representing
phase space and liquify his bag of stars so that the contents dribble
out at a constant rate.  Encumbered by his heavy bag of stellar light,
the monkey sits at one point $(\b x_i,\b v_i)$ dribbling starlight
unless disturbed.  Every so often, however, we squeeze his tail and he
jumps to a new position $(\b x_{i+1},\b v_{i+1})$ with probability
density~$\alpha(\b x_{i+1},\b v_{i+1})$.  Let the tail squeezes be a
Poisson process with rate~$\lambda$, so that the distribution of time
intervals between consecutive squeezes is $\pr(t)=\lambda\e^{-\lambda
  t}$.  More generally, the distribution of times between the $i^{\rm
  th}$ and $(i+r)^{\rm th}$ squeezes follows a gamma distribution.
Then, if the monkey lands $r$ times in a cell of volume~$\delta V$,
the total length of time he spends in that cell has probability
density
\begin{equation}
\pr(t|r) =
\begin{cases}
  \lambda\delta(\lambda t), & \text{$r=0$};\\
  \lambda\e^{-\lambda t}(\lambda t)^{r-1}/(r-1)!, & \text{$r>0.$}
\end{cases}
\end{equation}
In the limit of many tail
squeezes, the probability that he lands $r$ times in a cell of
volume~$\delta V$ is again given by~(\ref{eq:poisson}).
Summing over~$r$, the probability distribution for the length of time
he spends in the cell is
\begin{equation}
\begin{split}
\pr(t|\mu) & = \sum_{r=0}^\infty\pr(t|r)\pr(r|\mu)\\
& = 
\lambda\e^{-\mu}\left[\delta(\lambda t) + 
\e^{-\lambda t}\sqrt{\mu/\lambda t}\,I_1(2\sqrt{\lambda
  t\mu})\right],
\label{tmpprobt}
\end{split}
\end{equation}
with the Bessel function $I_1$ coming in through the identity
\begin{equation}
\sum_{r=0}^\infty {x^r\over r!(r+1)!} = {1\over\sqrt x}I_1(2\sqrt{x}).
\end{equation}
Equation~(\ref{tmpprobt}) is the ID prior~(\ref{eq:incontinent}) with
$F=\lambda t$.

\end{document}